# Scattering states solutions of Klein-Gordon equation with three physically solvable potential models

O.J. Oluwadare[*] and K.J. Oyewumi[+]

[*]Department of Physics, Federal University Oye-Ekiti, P. M. B. 373, Ekiti State, Nigeria
[+]Department of Physics, Federal University of Technology, Minna, Niger State, Nigeria.

## Abstract

The scattering state solutions of the Klein-Gordon equation with equal scalar and vector Varshni, Hellmann and Varshni-Shukla potentials for any arbitrary angular momentum quantum number $l$ are investigated within the framework of the functional analytical method using a suitable approximation. The asymptotic wave functions, approximate scattering phase shifts, normalization constants and bound state energy equations were obtained. The non-relativistic limits of the scattering phase shifts and the bound states energy equations for the three potentials were also obtained. Our bound states energy equations are in excellent agreement with the available ones in the literature. Our numerical and graphical results indicate the dependence of phase shifts on the screening parameter $\beta$, the potential parameter $b$ and angular momentum quantum number $l$.

## 1. INTRODUCTION

The solutions to relativistic equations are very essential in many aspects of modern physics, particularly, the Klein-Gordon equation is the most suitably used wave equation for the treatment of spinless particles in relativistic quantum mechanics. With various methods, Klein-Gordon equation has been studied with some exactly solvable potentials [1-9]. The scattering state solutions of various potential models have been a subject of considerable interest in the recent year because it gives an insight to understand atomic structures, electronic configuration of atoms, resonance and many collisions process [9-17].

However, the scattering state solutions of the Klein-Gordon equation with the Varshni, Hellman and Varshni-Shukla potentials needed to be studied due to the roles of these potential models in many aspect of physics. The Varshni potential is a short range repulsive potential energy function which has been used to describe the bound states of the interaction systems and has been applied in both classical and modern physics. Varshni potential is one of the potential energy functions that has been studied in the framework of the Schrödinger equation and it also plays a fundamental role in chemical and molecular Physics [18, 19].

The Varshni potential function was studied by Lim using the 2-body Kaxiras-Pandey parameters. In his work, he reported that Kaxiras and Pandey used this potential to describe the 2-

Corresponding author: oluwatimilehin.oluwadare@fuoye.edu.ng
[+]On Sabbatical Leave from: Theoretical Physics Section, Department of Physics, University of Ilorin, Ilorin, Nigeria. kjoyewumi66@unilorin.edu.ng

body energy portion of multi-body condensed matter [19].

The Hellmann potential [20-24] has been explored to explain the electron-core interaction [25], electron-ion [26], alkali hydride molecules and inner-shell ionization problem [27]. Very recently, Hamzavi et al. (2013) obtained the approximate solution of Hellmann potential within the framework of parametric Nikiforov-Uvarov method while the numerical result were confirmed using amplitude phase method [28]. Arda and Sever (2014) studied the approximate analytical solutions of the Dirac equation for Hellmann, Wei-Hua and Varshni potentials with their discussions centered on pseudospin and spin symmetric cases [29].

Onate et al. (2014) looked into the approximate eigen solutions of DKP and Klein-Gordon equations with Hellmann via super-symmetric approach. The solution of Schrödinger equation with Hellmann was obtained by formula methods. The behaviour of the energy in the first three states were presented [30]. Also, in 2015, Yazarloo et al. extended the study to scattering states of Dirac equation for spin and pseudospin symmetries with Hellmann potential [31].

The Varshni–Shukla potential [32, 33] has been used to obtain some spectroscopic constants [32] and also to evaluate some properties of Alkalis Chakogenide crystals in condensed matter physics [33]. Besides, Thakur and Mandal (2013) used this short range repulsive interaction to investigate the alkalis halide molecules [32].

Therefore, our aim is to study the scattering state solutions of Klein-Gordon equation with the equal scalar and vector Varshni, Hellmann and Varshni-Shukla potentials. The scattering state solutions are discussed via functional analytical method.

This paper is organized as follows; Section 2 contains the basic equations including the Klein-Gordon equation with the equal scalar and vector, Varshni, Hellmann and Varshni-Shukla potentials. In Section 3, we investigate the scattering state solutions of the Klein-Gordon equation with the equal scalar and vector Varshni, Hellmann and Varshni-Shukla potentials. Section 4 contains the numerical and graphical results and finally, the conclusion is given in Section 5.

## 2. The Basic Equations

The Klein–Gordon equation with the scalar potential $S(r)$ and vector potential $V(r)$ in the natural unit $\hbar = c = 1$ is given [9] as

$$\{-\nabla^2 + [M + S(r)]^2\}\psi(r,\theta,\varphi) = [E - V(r)]^2 \psi(r,\theta,\varphi), \qquad (1)$$

where $E$ is the relativistic energy of the system and $M$ is the rest mass of the spinless particles.

Using the Alhaidari condition [34] for evaluating negative energy states from positive energy states and applying $\psi_{nlm}(r,\theta,\varphi) = r^{-1}U_{nl}(r)Y_{lm}(\theta,\varphi)$ in Eq. (1), we obtain the time-independent radial Klein-Gordon equation for a spinless particles with the mixed scalar S(r) and vector V(r) potentials as:

$$\frac{d^2 U(r)}{dr^2} + \left\{ \left[\frac{1}{2}V(r) - E\right]^2 - \left[\frac{1}{2}S(r) - M\right]^2 - \frac{l(l+1)}{r^2} \right\} U(r) = 0. \tag{2}$$

The radial equation for spinless particles within equal scalar and vector potential, $V(r) = S(r)$ may now be written as:

$$\frac{d^2 U(r)}{dr^2} + \left[(E^2 - M^2) - (E+M)V(r) - \frac{l(l+1)}{r^2}\right] U(r) = 0. \tag{3}$$

Since the above radial equation has no exact solution, we apply an approximation

$$\frac{1}{r^2} \approx \frac{\beta^2}{(1-e^{-\beta r})^2} \; or \; \frac{1}{r} \approx \frac{\beta}{(1-e^{-\beta r})}, \tag{4}$$

throughout the study to deal with the centrifugal terms. This approximation has been reported to be valid for $\beta r \ll 1$ [35, 36].

### 3.0 Scattering State Solutions
### 3.1 Scattering State Solutions of Varshni potential (VP)

In this work, we consider the equal scalar and vector Varshni potentials [18, 19]

$$V_{VP}(r) = S_{VP}(r) = a\left[1 - \frac{b}{r}e^{-\beta r}\right], \tag{5}$$

where $r$ is the internuclear distance, $a$ and $b$ are the strengths of the potential and $\beta$ is the screening parameter which controls the shape of the potential energy curve.

With approximation in Eq. (4), the Klein-Gordon equation with Varshni potential may be transformed by the variable $z = 1 - e^{-\beta r}$, yielding

$$\frac{d^2 U_{VP}(z)}{dz^2} - \frac{1}{(1-z)}\frac{dU_{VP}(z)}{dz} + \frac{1}{z^2(1-z)^2}[P_{VP}z^2 + Q_{VP}z + R_{VP}]U_{VP}(z) = 0, \tag{6}$$

with

$$P_{VP} = \frac{k_{VP}^2}{\beta^2} + l(l+1) - \frac{a(E+M)b}{\beta}, \; Q_{VP} = \frac{a(E+M)b}{\beta}, \; R_{VP} = -l(l+1), \tag{7}$$

where $k_{VP} = \sqrt{(E^2 - M^2) - a(E+M) - l(l+1)\beta^2}$ is the asymptotic wave number of the spinless particles in this Varshni potential. Assuming the wave function of the form

$$U_{VP}(z) = z^{\lambda_{VP}}(1-z)^{-i(k_{VP}/\beta)}f(z), \tag{8}$$

and substituting it into Eq. (6), leads to the hypergeometric equation of the type [37]

$$z(1-z)f''(z) + \left[2\lambda_{VP} - \left(2\lambda_{VP} - 2i\frac{k_{VP}}{\beta} + 1\right)z\right]f'(z) + \left[\left(\lambda_{VP} - i\frac{k_{VP}}{\beta}\right)^2 + P_{VP}\right]f(z) = 0, \quad (9)$$

with the following phase shift parameters

$$\lambda_{VP} = \frac{1}{2} + \frac{1}{2}\sqrt{1 + 4l(l+1)} = l + 1, \quad (10)$$

$$\xi_1 = \lambda_{VP} - i\frac{k_{VP}}{\beta} - \sqrt{\frac{a(E+M)b}{\beta} - l(l+1) - \frac{k_{VP}^2}{\beta^2}}, \quad (11)$$

$$\xi_2 = \lambda_{VP} - i\frac{k_{VP}}{\beta} + \sqrt{\frac{a(E+M)b}{\beta} - l(l+1) - \frac{k_{VP}^2}{\beta^2}}, \quad (12)$$

$$\xi_3 = 2\lambda_{VP} = 2l + 2. \quad (13)$$

The radial wave functions for any arbitrary $l$−wave scattering states for the Varshni potential is obtained as:

$$U_{VP}(r) = N_{VP}(1 - e^{-\beta r})^{\lambda_{VP}} e^{ik_{VP}r} {}_2F_1(\xi_1, \xi_2, \xi_3; 1 - e^{-\beta r}), \quad (14)$$

where $N_{VP}$ is the normalization constant to be determined.

### 3.1.1 Bosonic Phase Shifts and Normalization Constant for the Varshni Potential

The phase shifts $\delta_{l,VP}$ and normalization constant $N_{VP}$ can be analyzed by applying the recurrence relation of hypergeometric function or analytic-continuation formula [37]

$${}_2F_1(\xi_1, \xi_2, \xi_3; z) = \frac{\Gamma(\xi_3)\Gamma(\xi_3 - \xi_1 - \xi_2)}{\Gamma(\xi_3 - \xi_1)\Gamma(\xi_3 - \xi_2)} {}_2F_1(\xi_1;\ \xi_2;\ 1 + \xi_1 + \xi_2 - \xi_3;\ 1 - z)$$

$$+ (1-z)^{\xi_3 - \xi_1 - \xi_2} \frac{\Gamma(\xi_3)\Gamma(\xi_1 + \xi_2 - \xi_3)}{\Gamma(\xi_1)\Gamma(\xi_2)} {}_2F_1(\xi_3 - \xi_1;\ \xi_3 - \xi_2;\ \xi_3 - \xi_1 - \xi_2 + 1;\ 1 - z). \quad (15)$$

Eq. (15) with the condition that ${}_2F_1(\xi_1, \xi_2, \xi_3; 0) = 1$, when $r \to \infty$, leads to

$${}_2F_1(\xi_1, \xi_2, \xi_3; 1 - e^{-\beta r}) \xrightarrow{r \to \infty} \Gamma(\xi_3)\left|\frac{\Gamma(\xi_3 - \xi_1 - \xi_2)}{\Gamma(\xi_3 - \xi_1)\Gamma(\xi_3 - \xi_2)} + e^{-2ik_{VP}r}\left|\frac{\Gamma(\xi_3 - \xi_1 - \xi_2)}{\Gamma(\xi_3 - \xi_1)\Gamma(\xi_3 - \xi_2)}\right|^*\right|, \quad (16)$$

where we have used the following phase shift relations:

$$\xi_3 - \xi_1 - \xi_2 = (\xi_1 + \xi_2 - \xi_3)^* = 2i(k_{VP}/\beta), \quad (17)$$

$$\xi_3 - \xi_2 = \lambda_{VP} + i\frac{k_{VP}}{\beta} - \sqrt{\frac{a(E+M)b}{\beta} - l(l+1) - \frac{k_{VP}^2}{\beta^2}} = \xi_1^*, \quad (18)$$

$$\xi_3 - \xi_1 = \lambda_{VP} + i\frac{k_{VP}}{\beta} + \sqrt{\frac{a(E+M)b}{\beta} - l(l+1) - \frac{k_{VP}^2}{\beta^2}} = \xi_2^*. \quad (19)$$

Now, by taking

$$\frac{\Gamma(\xi_3 - \xi_1 - \xi_2)}{\Gamma(\xi_3 - \xi_1)\Gamma(\xi_3 - \xi_2)} = \left|\frac{\Gamma(\xi_3 - \xi_1 - \xi_2)}{\Gamma(\xi_3 - \xi_1)\Gamma(\xi_3 - \xi_2)}\right| e^{i\Delta_l^{VP}}, \quad (20)$$

and substituting this into Eq. (16), we have

$${}_2F_1(\xi_1, \xi_2, \xi_3; 1 - e^{-\beta r}) \xrightarrow{r \to \infty} \Gamma(\xi_3)\left[\frac{\Gamma(\xi_3 - \xi_1 - \xi_2)}{\Gamma(\xi_3 - \xi_1)\Gamma(\xi_3 - \xi_2)}\right]$$

$$\times e^{-ik_{VP}r}\left[e^{i(k_{VP}r-\Delta_l^{VP})}+e^{-i(k_{VP}r-\Delta_l^{VP})}\right]. \tag{21}$$

Consequently, we obtain the asymptotic form of Eq. (14) for $r \to \infty$ as;

$$U_{VP}(r) \xrightarrow{r\to\infty} 2N_{l,VP}\Gamma(\xi_3)\left[\frac{\Gamma(\xi_3-\xi_1-\xi_2)}{\Gamma(\xi_3-\xi_1)\,\Gamma(\xi_3-\xi_2)}\right]\sin\left(k_{VP}r+\Delta_l^{VP}+\frac{\pi}{2}\right). \tag{22}$$

Finally, with the appropriate the boundary condition, Eq. (22) yields [38]

$$U_{VP}(\infty) \to 2\sin\left(k_{VP}r+\delta_{l,VP}-\frac{l\pi}{2}\right). \tag{23}$$

The phase shifts expression and the normalization constant are obtained respectively as follows:

$$\delta_{l,VP} = \frac{\pi}{2}+\frac{l\pi}{2}+\Delta_l^{VP}=\frac{\pi}{2}(l+1)+arg\Gamma(2i(k_{VP}/\beta))-arg\Gamma(\xi_2^*)-arg\Gamma(\xi_1^*) \tag{24}$$

and

$$N_{l,VP} = \frac{1}{\sqrt{\xi_3}}\left|\frac{\Gamma(\xi_1^*)\,\Gamma(\xi_2^*)}{\Gamma(2i(k_{VP}/\beta))}\right|. \tag{25}$$

### 3.1.2 Non-relativistic phase shifts for the Varshni potential

Here, we consider the non-relativistic (NR) limit [39-41] using a case when $-M \to E_{nl}$, $E+M \to \frac{2\mu}{\hbar^2}$ and $E^2-M^2 = \frac{2\mu E_{nl}}{\hbar^2}$, the non-relativistic phase shifts expression for the Varshni potential is obtained as:

$$\delta_{l,NR\ VP} = \frac{\pi}{2}(l+1)+arg\Gamma(2i(k_{NRVP}/\beta))-arg\Gamma(\xi_{NRVP2}^*)-arg\Gamma(\xi_{NRVP1}^*) \tag{26}$$

$$\xi_{NR1}^* = \lambda_{NRVP}+i\frac{k_{NRVP}}{\beta}-\sqrt{\frac{2\mu ab}{\hbar^2\beta}-l(l+1)-\frac{k_{NRVP}^2}{\beta^2}}. \tag{27}$$

$$\xi_{NR2}^* = \lambda_{NRVP}+i\frac{k_{NRVP}}{\beta}+\sqrt{\frac{2\mu ab}{\hbar^2\beta}-l(l+1)-\frac{k_{NRVP}^2}{\beta^2}}, \tag{28}$$

$$k_{NRVP} = \sqrt{\frac{2\mu(E-a)}{\hbar^2}-\beta^2 l(l+1)}\ ,\ \lambda_{NRVP} = l+1. \tag{29}$$

and

$$N_{l,NRVP} = \frac{1}{\sqrt{\xi_{NR3}}}\left|\frac{\Gamma(\xi_{NR1}^*)\,\Gamma(\xi_{NR2}^*)}{\Gamma(2i(k_{NRVP}/\beta))}\right| \tag{30}$$

where $\xi_{NR3} = \xi_3$.

### 3.1.3 Bosonic bound state energy equation for the Varshni Potential

In this sub-section, we investigate the analytical properties of partial-wave s-matrix to obtain the bound state energy at the poles of the s-matrix in the complex energy plane. And therefore, we need to discuss the Gamma function $\Gamma(\xi_3-\xi_1)$ [38] as:

$$\xi_3-\xi_1 = \lambda_{VP}+i\frac{k_{VP}}{\beta}+\sqrt{\frac{a(E+M)b}{\beta}-l(l+1)-\frac{k_{VP}^2}{\beta^2}}. \tag{31}$$

The first order poles of $\Gamma\left(\lambda_{VP} + i\frac{k_{VP}}{\beta} + \sqrt{\frac{a(E+M)b}{\beta} - l(l+1) - \frac{k_{VP}^2}{\beta^2}}\right)$ are situated at

$$\Gamma\left(\lambda_{VP} + i\frac{k_{VP}}{\beta} + \sqrt{\frac{a(E+M)b}{\beta} - l(l+1) - \frac{k_{VP}^2}{\beta^2}}\right) + n = 0 \quad (n = 0, 1, 2, \ldots). \quad (32)$$

Consequently, the relativistic bound state energy equation for the Varshni potential is obtained as

$$E^2 - M^2 - a(E+M) - \beta^2 l(l+1) + \beta^2 \left[\frac{(n+\lambda_{VP})^2 + l(l+1) - \frac{a(E+M)b}{\beta}}{2(n+\lambda_{VP})}\right]^2 = 0. \quad (33)$$

### 3.1.4 Non-relativistic bound state energy equation for the Varshni potential

By applying the same transformation variables in sub-section 3.1.2, the non-relativistic bound state energy for the Varshni potential is obtained as

$$E_{nl}^{NRVP} = a + \frac{\hbar^2 \beta^2 l(l+1)}{2\mu} - \frac{\hbar^2 \beta^2}{2\mu}\left[\frac{(n+l+1)^2 + l(l+1) - \frac{2\mu ab}{\hbar^2 \beta}}{2(n+l+1)}\right]^2. \quad (34)$$

### 3.2 Scattering State Solutions of Hellmann potential (HP)

In this work, we consider the equal scalar and vector Hellman potential [20-24]

$$V_{HP}(r) = S_{HP}(r) = -\frac{a}{r} + \frac{b}{r}e^{-\beta r}, \quad (35)$$

which is a superposition of the Coulomb and Yukawa potential with $r$ as the internuclear distance, $a$ and $b$ are the strengths of the Coulomb and Yukawa potential respectively and $\beta$ is the screening parameter which controls the shape of the potential energy curve.

Using the same approximation in Eq. (4), the Klein-Gordon equation with Hellman potential may be transformed by the variable $z = 1 - e^{-\beta r}$, turning to the differential equation

$$\frac{d^2 U_{HP}(z)}{dz^2} - \frac{1}{(1-z)}\frac{dU_{HP}(z)}{dz} + \frac{1}{z^2(1-z)^2}[P_{HP}z^2 + Q_{HP}z + R_{HP}]U_{HP}(z) = 0, \quad (36)$$

with

$$P_{HP} = \frac{k_{HP}^2}{\beta^2} + l(l+1) - \frac{a(E+M)}{\beta} - \frac{b(E+M)}{\beta}, \quad Q_{HP} = \frac{(E+M)(a-b)}{\beta}, \quad R_{HP} = -l(l+1), \quad (37)$$

where $k_{HP} = \sqrt{(E^2 - M^2) + a(E+M)\beta - l(l+1)\beta^2}$ is the asymptotic wave number of the spinless particle in Hellmann potential.

### 3.2.1 Bosonic Phase Shifts and Normalization Constant for the Hellmann Potential

To avoid repetition in the method of analysis, we employ the previous procedure in subsection 3.1 and obtain the phase shifts for the Hellmann potential as

$$\delta_{l,HP} = \frac{\pi}{2}(l+1) + arg\Gamma(2i(k_{HP}/\beta)) - arg\Gamma(\eta_2{}^*) - arg\Gamma(\eta_1{}^*), \qquad (38)$$

with the following phase shift parameters

$$\eta_2{}^* = \lambda_{HP} + i\frac{k_{HP}}{\beta} + \sqrt{\frac{a(E+M)}{\beta} - \frac{b(E+M)}{\beta} - l(l+1) - \frac{k_{HP}^2}{\beta^2}}, \qquad (39)$$

$$\eta_1{}^* = \lambda_{HP} + i\frac{k_{HP}}{\beta} - \sqrt{\frac{a(E+M)}{\beta} - \frac{b(E+M)}{\beta} - l(l+1) - \frac{k_{HP}^2}{\beta^2}}, \qquad (40)$$

and the normalization constant

$$N_{l,HP} = \frac{1}{\sqrt{\eta_3}}\left|\frac{\Gamma(\eta_1{}^*)\,\Gamma(\eta_2{}^*)}{\Gamma(2i(k_{HP}/\beta))}\right|, \qquad \eta_3 = 2\lambda_{HP} = 2l+2. \qquad (41)$$

The relativistic radial wave function for the scattering of the Hellmann potential is obtained as:

$$U_{HP}(r) = N_{l,HP}(1-e^{-\beta r})^{\lambda_{HP}} e^{ik_{HP}r} {}_2F_1(\eta_1,\eta_2,\eta_3; 1-e^{-\beta r}), \qquad (42)$$

with the following wave functions parameters

$$\eta_1 = \lambda_{HP} - i\frac{k_{HP}}{\beta} - \sqrt{\frac{a(E+M)}{\beta} - \frac{b(E+M)}{\beta} - l(l+1) - \frac{k_{HP}^2}{\beta^2}}, \qquad (43)$$

$$\eta_2 = \lambda_{HP} - i\frac{k_{HP}}{\beta} + \sqrt{\frac{a(E+M)}{\beta} - \frac{b(E+M)}{\beta} - l(l+1) - \frac{k_{HP}^2}{\beta^2}}, \qquad (44)$$

$$\lambda_{HP} = \frac{1}{2} + \frac{1}{2}\sqrt{1+4l(l+1)} = l+1. \qquad (45)$$

### 3.2.2 Non-relativistic phase shift for the Hellmann potential

In a similar fashion, the non-relativistic phase shifts expression for the Hellmann potential is obtained as:

$$\delta_{l,NRHP} = \frac{\pi}{2}(l+1) + arg\Gamma(2i(k_{NRHP}/\beta)) - arg\Gamma(\eta_{NRHP2}{}^*) - arg\Gamma(\eta_{NRHP1}{}^*), \qquad (46)$$

$$\eta_{NR1}{}^* = l+1 + i\frac{k_{NRHP}}{\beta} - \sqrt{\frac{2\mu a}{\hbar^2\beta} - l(l+1) - \frac{2\mu b}{\hbar^2\beta} - \frac{k_{NRHP}^2}{\beta^2}}, \qquad (47)$$

$$\eta_{NR2}{}^* = l+1 + i\frac{k_{NRHP}}{\beta} + \sqrt{\frac{2\mu a}{\hbar^2\beta} - l(l+1) - \frac{2\mu b}{\hbar^2\beta} - \frac{k_{NRHP}^2}{\beta^2}}, \qquad (48)$$

$$k_{NRHP} = \sqrt{\frac{2\mu E}{\hbar^2} + \frac{2\mu a\beta}{\hbar^2} - \beta^2 l(l+1)}. \qquad (49)$$

### 3.2.3 Bosonic Bound State Energy Equation for the Hellmann Potential

In the same way, we need to discuss the Gamma function $\Gamma(\eta_3 - \eta_1)$ [38] as:

$$\eta_3 - \eta_1 = \lambda_{HP} + i\frac{k_{HP}}{\beta} + \sqrt{\frac{a(E+M)}{\beta} - \frac{b(E+M)}{\beta} - l(l+1) - \frac{k_{HP}^2}{\beta^2}}. \qquad (50)$$

The first order poles of $\Gamma\left(\lambda_{HP} + i\frac{k_{HP}}{\beta} + \sqrt{\frac{a(E+M)}{\beta} - \frac{b(E+M)}{\beta} - l(l+1) - \frac{k_{HP}^2}{\beta^2}}\right)$ are situated at

$$\Gamma\left(\lambda_{HP} + i\frac{k_{HP}}{\beta} + \sqrt{\frac{a(E+M)}{\beta} - \frac{b(E+M)}{\beta} - l(l+1) - \frac{k_{HP}^2}{\beta^2}}\right) + n = 0 \quad (n = 0, 1, 2, \ldots). \quad (51)$$

Consequently, the bosonic energy equation for the Hellmann potential is obtained as

$$E^2 - M^2 + a\beta(E+M) - \beta^2 l(l+1) + \beta^2 \left[\frac{(n+\lambda_{HP})^2 + l(l+1) - \frac{a(E+M)}{\beta} + \frac{b(E+M)}{\beta}}{2(n+\lambda_{HP})}\right]^2 = 0. \quad (52)$$

### 3.2.4 Non-relativistic bound state energy equation for the Hellmann potential

By using the same transformation variables in sub-section 3.1.2, the non-relativistic bound state energy for the Hellmann potential is obtained as

$$E_{nl}^{NRHP} = \frac{\hbar^2 \beta^2 l(l+1)}{2\mu} - a\beta - \frac{\hbar^2 \beta^2}{2\mu}\left[\frac{(n+l+1)^2 + \frac{2\mu b}{\hbar^2 \beta} - \frac{2\mu a}{\hbar^2 \beta} + l(l+1)}{2(n+l+1)}\right]^2 \quad (53)$$

The bound states energy equation of eq. (53) is the same as the one obtained in Eq. (21) by Hamzavi et al. (2014) and Eq. (24) of Onate et al. 2014. See references [28] and [30] for confirmation of the results obtained.

### 3.3 Scattering state solution of Varshni-Shukla potential (VSP)

In this section, we consider the equal scalar and vector Varshni-Shukla potential [29, 30]

$$V_{VSP}(r) = S_{VSP}(r) = \frac{b}{r^2}e^{-\beta r}, \quad (54)$$

where $\beta = \frac{1}{\rho}$ and $\rho$ is the repulsive potential parameter. $V_{VSP}(r)$ diverges for small $r$. With the aid of the above approximation in Eq. (4), the Klein-Gordon equation with Varshni-Shukla potential may be transformed by the variable $z = 1 - e^{-\beta r}$ to obtain the differential equation

$$\frac{d^2 U_{VSP}(z)}{dz^2} - \frac{1}{(1-z)}\frac{dU_{VSP}(z)}{dz} + \frac{1}{z^2(1-z)^2}[P_{VSP}z^2 + Q_{VSP}z + R_{VSP}]U_{VSP}(z) = 0, \quad (55)$$

where

$$P_{VSP} = \frac{k_{VSP}^2}{\beta^2} + l(l+1), \quad Q_{VSP} = (E+M)b, \quad R_{VSP} = -[(E+M)b + l(l+1)], \quad (56)$$

where $k_{VSP} = \sqrt{(E^2 - M^2) - l(l+1)\beta^2}$ is the asymptotic wave number of the bosonic particle on the Varshni-Shukla (VSP) potential.

### 3.3.1 Phase Shift and Normalization Constant for the Varshni-Shukla Potential

Similarly, we follow the same procedure in subsection 3.1 and obtain the phase shifts for the Varshni-Shukla potential as

$$\delta_{l,VSP} = \frac{\pi}{2}(l+1) + arg\Gamma(2i(k_{VSP}/\beta)) - arg\Gamma(\zeta_2^*) - arg\Gamma(\zeta_1^*), \quad (57)$$

with the following phase shift parameters

$$\zeta_2^* = \lambda_{VSP} + i\frac{k_{VSP}}{\beta} + i\sqrt{l(l+1) + \frac{k_{VSP}^2}{\beta^2}}, \tag{58}$$

$$\zeta_1^* = \lambda_{VSP} + i\frac{k_{VSP}}{\beta} - i\sqrt{l(l+1) + \frac{k_{VSP}^2}{\beta^2}}, \tag{59}$$

and the corresponding normalization constant

$$N_{l,VSP} = \frac{1}{\sqrt{\zeta_3}} \left| \frac{\Gamma(\zeta_1^*)\Gamma(\zeta_2^*)}{\Gamma(2i(k_{VSP}/\beta))} \right|, \tag{60}$$

where

$$\zeta_3 = 2\lambda_{VSP}. \tag{61}$$

Consequently, the radial wave functions for the scattering on the Varshni-Shukla potential is obtained as:

$$U_{VSP}(r) = N_{l,VSP}(1 - e^{-\beta r})^{\lambda_{VSP}} e^{ik_{VSP}r} {}_2F_1(\zeta_1, \zeta_2, \zeta_3; 1 - e^{-\beta r}), \tag{62}$$

with the wave functions parameters

$$\zeta_1 = \lambda_{VSP} - i\frac{k_{VSP}}{\beta} - i\sqrt{l(l+1) + \frac{k_{VSP}^2}{\beta^2}}, \tag{63}$$

$$\zeta_2 = \lambda_{VSP} - i\frac{k_{VSP}}{\beta} + i\sqrt{l(l+1) + \frac{k_{VSP}^2}{\beta^2}}, \tag{64}$$

$$\lambda_{VSP} = \frac{1}{2} + \sqrt{\frac{1}{4} + l(l+1) + (E+M)b}. \tag{65}$$

### 3.3.2 Non-relativistic phase shifts for the Varshni-Shukla potential

In a similar way, the non-relativistic (NR) phase shifts expression for the Varshni-Shukla potential is obtained as:

$$\delta_{l,NRVSP} = \frac{\pi}{2}(l+1) + arg\Gamma(2i(k_{NRVSP}/\beta)) - arg\Gamma(\zeta_{NRVSP2}^*) - arg\Gamma(\zeta_{NRVSP1}^*). \tag{66}$$

$$\zeta_{NRVSP2}^* = \lambda_{NRVSP} + i\frac{k_{NRVSP}}{\beta} + i\sqrt{l(l+1) + \frac{k_{NRVSP}^2}{\beta^2}}, \tag{67}$$

$$\zeta_{NRVSP1}^* = \lambda_{NRVSP} + i\frac{k_{NRVSP}}{\beta} - i\sqrt{l(l+1) + \frac{k_{NRVSP}^2}{\beta^2}}, \tag{68}$$

$$k_{NRVSP} = \sqrt{\frac{2\mu E}{\hbar^2} - \beta^2 l(l+1)}, \tag{69}$$

$$\lambda_{NRVSP} = \frac{1}{2} + \sqrt{\frac{1}{4} + l(l+1) + \frac{2\mu b}{\hbar^2}}. \tag{70}$$

### 3.3.3 Bosonic Energy equation for the Varshni-Shukla Potential

In the same way, we need to analyze the Gamma function $\Gamma(\zeta_2^*)$[38] as:

$$\zeta_2^* = \lambda_{VSP} + i\frac{k_{VSP}}{\beta} + i\sqrt{l(l+1) + \frac{k_{VSP}^2}{\beta^2}}. \tag{71}$$

The first order poles of $\Gamma\left(\lambda_{VSP} + i\frac{k_{VSP}}{\beta} + i\sqrt{l(l+1) + \frac{k_{VSP}^2}{\beta^2}}\right)$ are situated at

$$\Gamma\left(\lambda_{VSP} + i\frac{k_{VSP}}{\beta} + i\sqrt{l(l+1) + \frac{k_{VSP}^2}{\beta^2}}\right) + n = 0 \quad (n = 0, 1, 2, \dots). \tag{72}$$

Therefore, the bosonic energy equation for the Varshni-Shukla potential is obtained as

$$E^2 - M^2 - \beta^2 l(l+1) + \beta^2 \left[\frac{(n+\lambda_{VSP})^2 + l(l+1)}{2(n+\lambda_{VSP})}\right]^2 = 0. \tag{73}$$

### 3.3.4 Non-relativistic bound state energy spectral for the Varshni-Shukla potential

By using the same transformation variables in sub-section 3.1.2, the non-relativistic bound state energy for the Varshni-Shukla potential is obtained as

$$E_{nl}^{NRVSP} = \frac{\hbar^2 \beta^2 l(l+1)}{2\mu} - \frac{\hbar^2 \beta^2}{2\mu}\left[\frac{\left(n+\frac{1}{2}+\sqrt{\frac{1}{4}+l(l+1)+\frac{2\mu b}{\hbar^2}}\right)^2 + l(l+1)}{2\left(n+\frac{1}{2}+\sqrt{\frac{1}{4}+l(l+1)+\frac{2\mu b}{\hbar^2}}\right)}\right]^2. \tag{74}$$

## 4.0. Numerical Results and Discussions

Table 1: Bosonic Phase shifts for the Varshni, Hellmann and Varshni-Shukla potentials with $a = 2$ and $E = M = b = 1$.

| $l$ | $\beta$ | $\delta_{l,VP}$ | $\delta_{l,HP}$ | $\delta_{l,VSP}$ |
|---|---|---|---|---|
| 0 | 0.2 | -56.42013 | -18.28023 | 1.57080 |
|   | 0.4 | -13.97716 | -8.70939 | 1.57080 |
|   | 0.6 | -3.67443 | -5.10888 | 1.57080 |
|   | 0.8 | 0.29281 | -3.19714 | 1.57080 |
|   | 1.0 | 2.15222 | -2.00889 | 1.57080 |
| 1 | 0.2 | -59.03413 | -17.45706 | 0.01257 |
|   | 0.4 | -17.09669 | -7.33489 | 0.01257 |
|   | 0.6 | -7.10043 | -3.38560 | 0.01257 |
|   | 0.8 | -3.31511 | -1.21715 | 0.01257 |
|   | 1.0 | -1.55237 | 0.17432 | 0.01257 |
| 2 | 0.2 | -62.66555 | -14.10380 | -4.62092 |
|   | 0.4 | -21.63335 | -2.49275 | -4.62092 |
|   | 0.6 | -12.09103 | 2.95692 | -4.62092 |
|   | 0.8 | -8.51103 | 4.58274 | -4.62092 |
|   | 1.0 | -6.82200 | 3.84423 | 1.05839 |
| 3 | 0.2 | -67.26676 | -7.47923 | -10.98016 |
|   | 0.4 | -27.38963 | 5.41503 | -10.98016 |
|   | 0.6 | -18.31684 | 3.06727 | -10.98016 |
|   | 0.8 | -14.91997 | 1.98887 | -10.98016 |
|   | 1.0 | -13.30400 | 1.36681 | -0.97221 |

Table 2: Bosonic Phase shifts for the Varshni and Hellmann potentials with $a = 2, \beta = 0.2$ and $E = M = 1$

| $l$ | $b$ | $\delta_{l,VP}$ | $\delta_{l,HP}$ |
|---|---|---|---|
| 0 | -2 | -61.22712 | -11.93829 |
|   | -1 | -59.56242 | -13.77770 |
|   | 0 | -57.96276 | -15.84454 |
|   | 1 | -56.42013 | -18.28023 |
|   | 2 | -54.92814 | -21.28685 |
| 1 | -2 | -63.77592 | -11.44881 |
|   | -1 | -62.13658 | -13.27676 |
|   | 0 | -60.55831 | -15.28066 |
|   | 1 | -59.03413 | -17.45706 |
|   | 2 | -57.55828 | -19.58049 |
| 2 | -2 | -67.28155 | -8.68290 |
|   | -1 | -65.69075 | -10.48117 |
|   | 0 | -64.15387 | -12.33642 |
|   | 1 | -62.66555 | -14.10380 |
|   | 2 | -61.22121 | -15.59596 |
| 3 | -2 | -71.70483 | -2.99836 |
|   | -1 | -70.18205 | -4.71752 |
|   | 0 | -68.70381 | -6.25399 |
|   | 1 | -67.26676 | -7.47923 |
|   | 2 | -65.86776 | -8.41824 |

Table 3: Bosonic Phase shifts for the Varshni, Hellmann and Varshni-Shukla potentials with $a = 0, \beta = 0.2$ and $E = M = 1$

| $l$ | $b$ | $\delta_{l,VP}$ | $\delta_{l,HP}$ | $\delta_{l,VSP}$ |
|---|---|---|---|---|
| 0 | -2 | 1.57080 | 14.13717 | 2.84043 |
|   | -1 | 1.57080 | 10.99557 | 3.41057 |
|   | 0 | (1.57080) | (1.57080) | (1.57080) |
|   | 1 | 1.57080 | 1.57080 | 1.57080 |
|   | 2 | 1.57080 | 1.57080 | 1.57080 |
| 1 | -2 | 0.76042 | 7.47286 | 2.20500 |
|   | -1 | 0.76042 | 4.42150 | 2.67105 |
|   | 0 | (0.76042) | (0.76042) | (0.76042) |
|   | 1 | 0.76042 | 2.27344 | 0.01257 |
|   | 2 | 0.76042 | 2.27344 | -0.47054 |
| 2 | -2 | -4.07243 | 0.31859 | -2.30028 |
|   | -1 | -4.07243 | -1.89586 | -3.35509 |
|   | 0 | (-4.07243) | (-4.07243) | (-4.07243) |
|   | 1 | -4.07243 | 1.05839 | -4.62092 |
|   | 2 | -4.07243 | 1.05839 | -5.06634 |
| 3 | -2 | -10.56258 | -7.45813 | -9.52555 |
|   | -1 | -10.56258 | -9.05505 | -10.08485 |
|   | 0 | (-10.56258) | (-10.56258) | (-10.56258) |
|   | 1 | -10.56258 | -0.97221 | -10.98016 |
|   | 2 | -10.56258 | -0.97221 | -11.35140 |

Table 4: Non-relativistic Phase shifts for the Varshni, Hellmann and Varshni-Shukla potentials with $a = 2$ and $E = \mu = \hbar = b = 1$.

| $l$ | $\beta$ | $\delta_{l,NRVP}$ | $\delta_{l,NRHP}$ | $\delta_{l,NRVSP}$ |
|---|---|---|---|---|
| 0 | 0.2 | -29.25966 | -58.79700 | -48.13367 |
|   | 0.4 | -4.19045 | -22.19149 | -14.92677 |
|   | 0.6 | 1.42510 | -11.95690 | -6.51696 |
|   | 0.8 | 3.38203 | -7.36428 | -3.10662 |
|   | 1.0 | 4.17896 | -4.81609 | -1.39647 |
| 1 | 0.2 | -32.12819 | -58.22093 | -46.13639 |
|   | 0.4 | -7.61506 | -21.05682 | -12.16825 |
|   | 0.6 | -2.28690 | -10.44818 | -3.15922 |
|   | 0.8 | -0.45903 | -5.57535 | 0.78843 |
|   | 1.0 | 0.30661 | -2.80362 | 3.14159 |
| 2 | 0.2 | -36.23320 | -56.05511 | -43.01706 |
|   | 0.4 | -12.61881 | -17.55281 | -6.96689 |
|   | 0.6 | -7.61456 | -5.84922 | 4.94071 |
|   | 0.8 | -5.86275 | 0.13296 | 4.20160 |
|   | 1.0 | -5.07798 | 4.71239 | 2.40269 |
| 3 | 0.2 | -41.46929 | -52.29246 | -38.27221 |
|   | 0.4 | -18.86863 | -11.08846 | 4.15755 |
|   | 0.6 | -14.13656 | 5.15002 | 2.34222 |
|   | 0.8 | -12.45038 | 4.06066 | 0.83600 |
|   | 1.0 | -11.67725 | 2.62919 | 0.17027 |

Table 5: Non-relativistic Phase shifts for the Varshni, Hellmann and Varshni-Shukla with $a = 2, \beta = 0.2$ and $E = \mu = \hbar = b = 1$

| l | b | $\delta_{l,NRVP}$ | $\delta_{l,NRHP}$ |
|---|---|---|---|
| 0 | -2 | -36.23205 | -53.86816 |
|   | -1 | -33.71126 | -55.34788 |
|   | 0  | -31.40448 | -56.97214 |
|   | 1  | -29.25966 | -58.79700 |
|   | 2  | -27.24356 | -60.81170 |
| 1 | -2 | -38.89140 | -53.90060 |
|   | -1 | -36.46611 | -55.28560 |
|   | 0  | -34.22453 | -56.73434 |
|   | 1  | -32.12818 | -58.22093 |
|   | 2  | -30.15019 | -59.68508 |
| 2 | -2 | -42.61406 | -52.36346 |
|   | -1 | -40.35832 | -53.60324 |
|   | 0  | -38.23766 | -54.84292 |
|   | 1  | -36.23320 | -56.05511 |
|   | 2  | -34.32834 | -57.21097 |
| 3 | -2 | -47.35474 | -49.17259 |
|   | -1 | -45.30924 | -50.25108 |
|   | 0  | -43.34815 | -51.29595 |
|   | 1  | -41.46929 | -52.29246 |
|   | 2  | -39.66651 | -53.23022 |

Table 6: Non-relativistic Phase shifts for the Varshni, Hellmann and Varshni-Shukla potentials with $a = 0, \beta = 0.2$ and $E = \mu = \hbar = b = 1$

| l | b | $\delta_{l,NRVP}$ | $\delta_{l,NRHP}$ | $\delta_{l,NRVSP}$ |
|---|---|---|---|---|
| 0 | -2 | -46.63347 | -42.38521 | -50.47872 |
|   | -1 | -46.63347 | -44.37636 | -48.50130 |
|   | 0  | (-46.63347) | (-46.63347) | (-46.63347) |
|   | 1  | -46.63347 | -48.72610 | -48.13366 |
|   | 2  | -46.63347 | -50.36426 | -48.94539 |
| 1 | -2 | -45.36959 | -42.09189 | -47.58161 |
|   | -1 | -45.36959 | -43.74554 | -44.01201 |
|   | 0  | (-45.36959) | (-45.36959) | (-45.36959) |
|   | 1  | -45.36959 | -46.85068 | -46.13639 |
|   | 2  | -45.36959 | -48.13969 | -46.72717 |
| 2 | -2 | -42.56397 | -39.96959 | -41.35359 |
|   | -1 | -42.56397 | -41.30953 | -42.02978 |
|   | 0  | (-42.56397) | (-42.56397) | (-42.56397) |
|   | 1  | -42.56397 | -43.70317 | -43.01706 |
|   | 2  | -42.56397 | -44.72157 | -43.41567 |
| 3 | -2 | -37.98095 | -35.91976 | -37.31654 |
|   | -1 | -37.98095 | -36.99512 | -37.66459 |
|   | 0  | (-37.98095) | (-37.98095) | (-37.98095) |
|   | 1  | -37.98095 | -38.87475 | -38.27222 |
|   | 2  | -37.98095 | -39.68215 | -38.54296 |

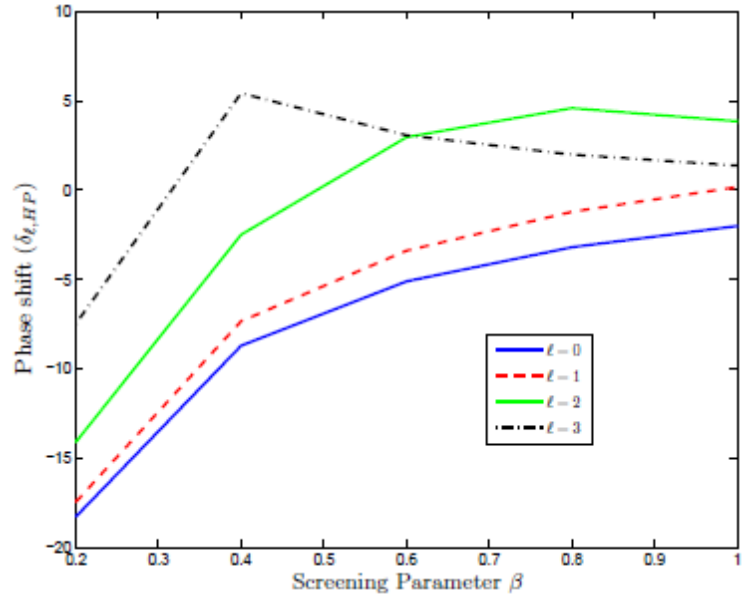

Figure 1 (a) Variation of bosonic phase shifts for the Hellmann potential as a function of screening parameter $\beta$ for $a = 2$ and $E = M = b = 1$.

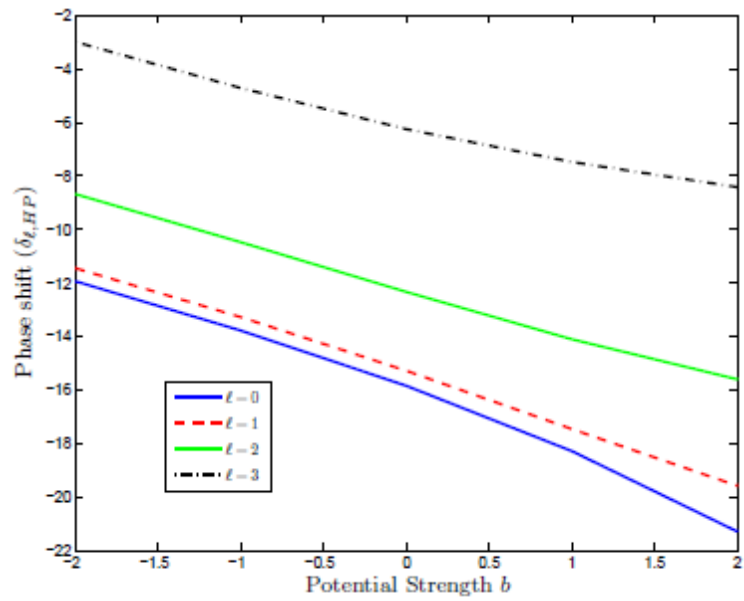

Figure 1 (b) Variation of bosonic phase shifts for the Hellmann potential as a function of potential strength $b$ for $a = 2$, $\beta = 0.2$ and $E = M = 1$.

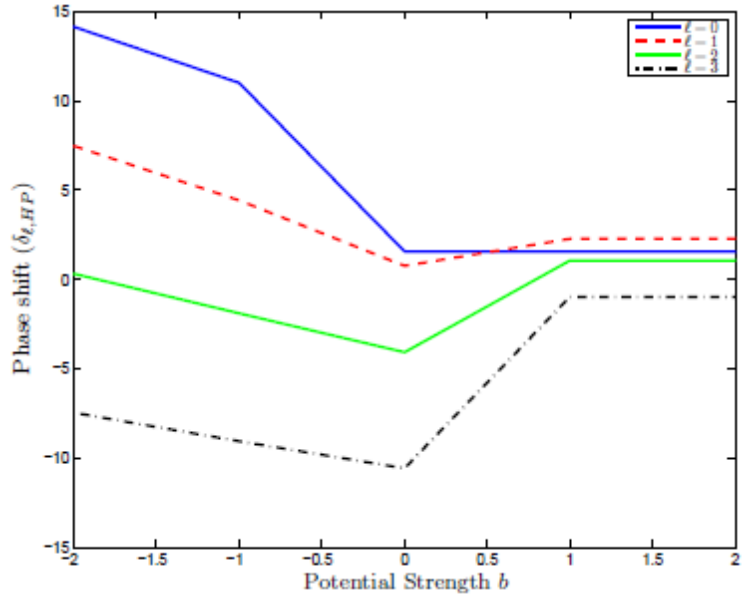

Figure 1 (c) Variation of bosonic phase shifts for the Hellmann potential as a function of potential strength $b$ for $a = 0$, $\beta = 0.2$ and $E = M = 1$

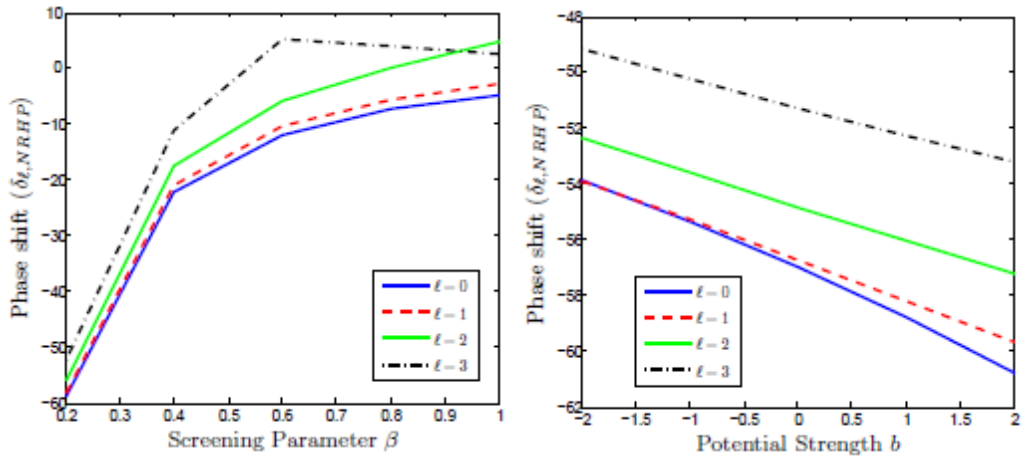

Figure 1 (d)                                                                    Figure 1 (e)

Figure 1 (d) Variation of non-relativistic phase shifts for the Hellmann potential as a function of screening parameter $\beta$ for $a = 2$ and $E = M = b = 1$

Figure 1 (e) Variation of non-relativistic phase shifts for the Hellmann potential as a function of potential strength $b$ for $= 2$, $\beta = 0.2$ and $E = M = 1$.

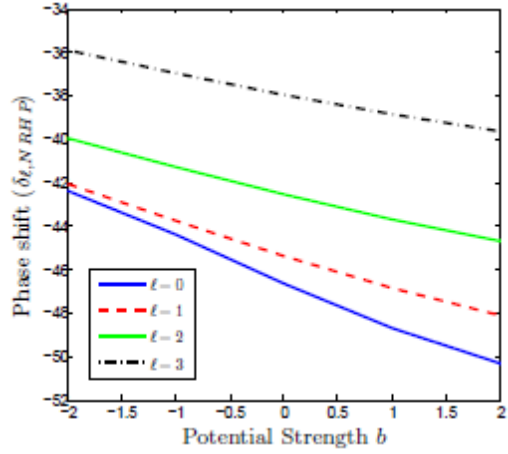

Figure 1 (f) Variation of non-relativistic phase shifts for the Hellmann potential as a function of potential strength $b$ for $a = 0$, $\beta = 0.2$ and $E = M = 1$

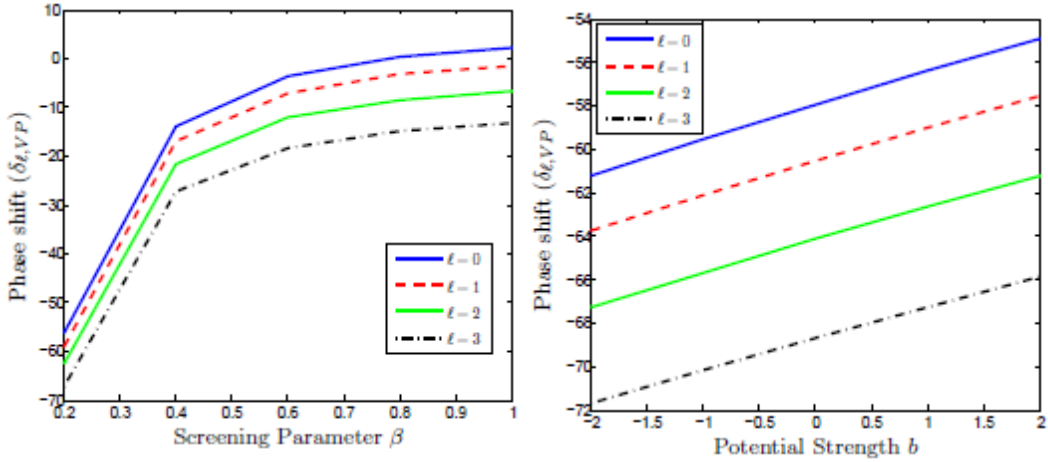

Figure 2 (a)            Figure 2 (b)

Figure 2 (a) Variation of bosonic phase shifts for the Varshni potential as a function of screening parameter $\beta$ for $a = 2$ and $E = M = b = 1$.

Figure 2 (b) Variation of bosonic phase shifts for the Varshni potential as a function of potential strength $b$ for $a = 2$, $\beta = 0.2$ and $E = M = 1$

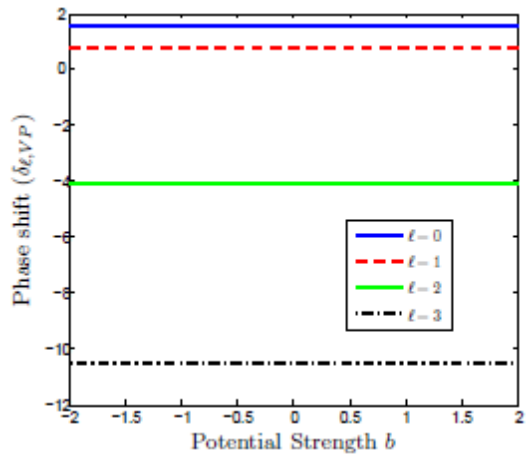

Figure 2 (c) Variation of bosonic phase shifts for the Varshni potential as a function of potential strength $b$ for a= 0, $\beta = 0.2$ and $E = M = 1$

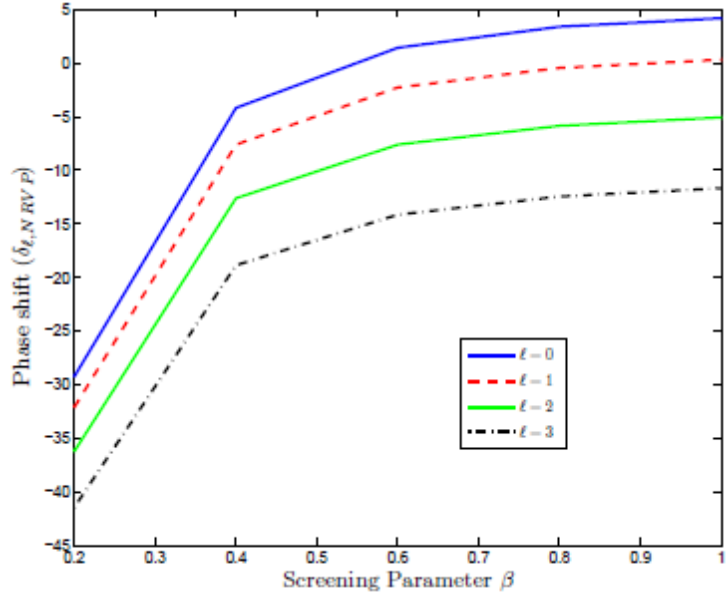

Figure 2 (d) Variation of non-relativistic phase shifts for the Varshni potential as a function of screening parameter $\beta$ for $a = 2$ and $= M = b = 1$.

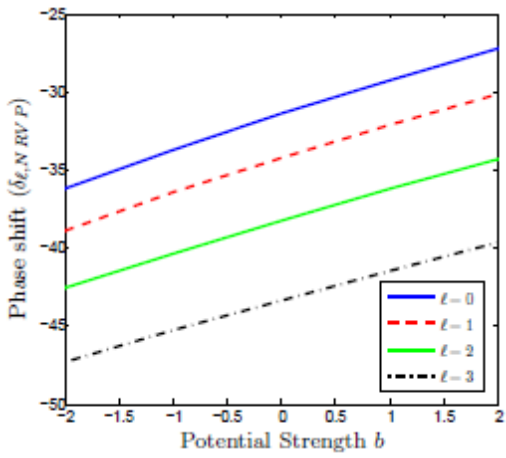 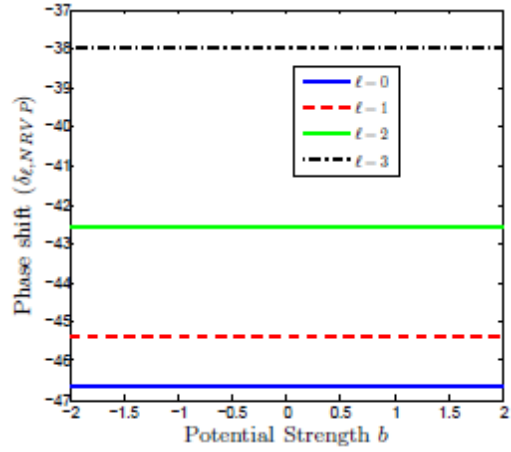

Figure 2 (e)                      Figure 2 (f)

Figure 2 (e) Variation of non-relativistic phase shifts for the Varshni potential as a function of potential strength $b$ for $= 2$, $\beta = 0.2$ and $E = M = 1$.

Figure 2 (f) Variation of non-relativistic phase shifts for the Varshni potential as a function of potential strength $b$ for $a = 0$, $\beta = 0.2$ and $E = M = 1$.

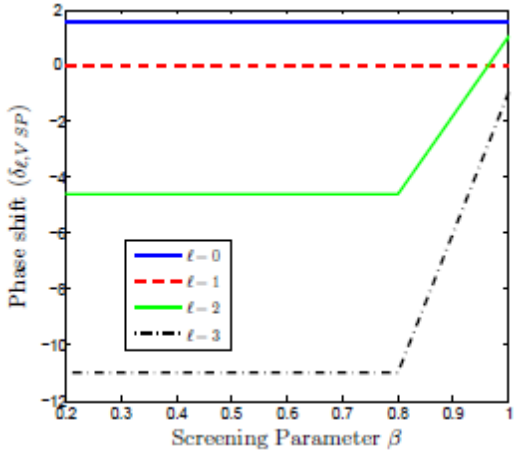 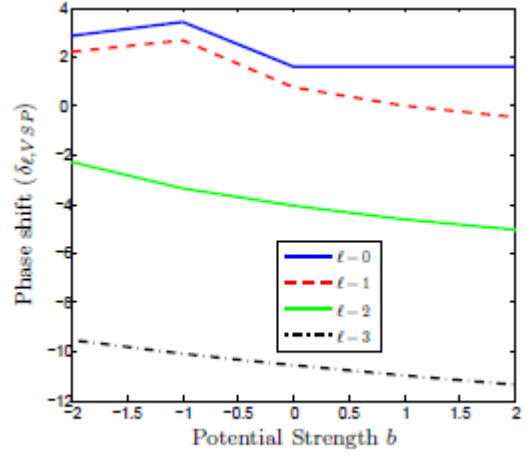

Figure 3 (a)                      Figure 3 (b)

Figure 3 (a) Variation of bosonic phase shifts for the Varshni-Shukla potential as a function of screening parameter $\beta$ for $a = 2$ and $= M = b = 1$.

Figure 3 (b) Variation of bosonic phase shifts for the Varshni-Shukla potential as a function of potential strength $b$ for, $\beta = 0.2$ and $E = M = 1$.

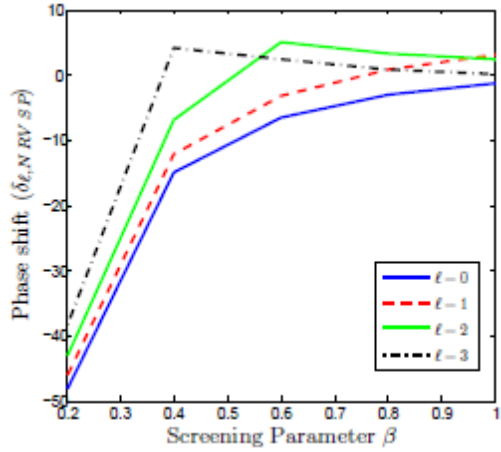 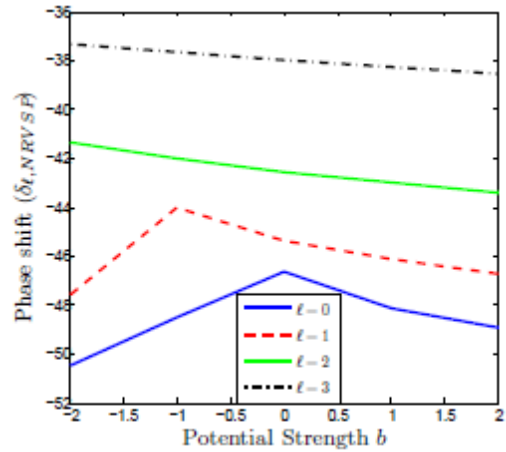

Figure 3 (c)                      Figure 3 (d)

Figure 3 (c) Variation of non-relativistic phase shifts for the Varshni-Shukla potential as a function of screening parameter $\beta$ for $a = 2$ and $= M = b = 1$.

Figure 3 (d) Variation of non-relativistic phase shifts for the Varshni-Shukla potential as a function of potential strength $b$ for, $\beta = 0.2$ and $E = M = 1$

To see the beauty of our work, we calculate the scattering phase shifts and made some meaningful plots. Table 1, figures 1(a), 2 (a) and 3 (a) illustrate the variation of bosonic phase shifts as a function of screening parameter $\beta$. Surprisingly, both Hellmann and Varshni potentials exhibit almost the same phase shift characteristic whereas Varshni-Shukla potential displayed different behaviour entirely. In this case, phase shift increases in the direction of increasing screening parameter $\beta$ but decreasing with increasing angular momentum number $l$.

Table 4, Figures 1(d), 2 (d) and 3 (c) illustrate the variation of non-relativistic phase shifts as a function of screening parameter $\beta$. The three potentials exhibit almost the same phase shift characteristic for the same potential parameters. The phase shifts rise exponentially with increasing screening parameter $\beta$ but decrease with increasing angular momentum number $l$.

Tables 2 and 3 illustrate the variation of bosonic phase shifts as a function of potential parameter $b$. In Table 2, Phase shift reduces with increasing angular momentum number $l$ for all the potentials. Phase shift reduces with increasing $b$ and increases with an increasing $b$ for Hellmann potential and Varshni potential respectively. In $\delta_{l,HP}$ column, the results of where $b = 0$ corresponds to that of attractive Coulomb potential.

In Table 3, the phase shift is constant for arbitrary values of the screening parameters $b$ for Varshni potential. It is also constant in the direction of positively increasing $b$ for Varshni and Hellmann potentials but increasing in the direction towards negatively increasing $b$. The result of when $a = 0$ and $b = 0$ corresponds to potentials free regions. The clarity of dependence of bosonic phase shifts on potential parameter $b$ are seen in figures 1 (b), 1 (c), 2 (b) and 2 (c).

Tables 5 and 6 illustrate the variation of non-relativistic phase shifts as a function of potential parameter $b$ for the three potentials. Both Varshni and Hellmann potentials displayed opposite characteristics in table 5. In Table 6, for any arbitrary angular momentum number $l$ and $b = 0$, the three potentials exhibit the same scattering phase shifts. This points correspond to potential free region. The dependence of non-relativistic phase shift on potential parameter $b$ are also seen in the figures 1 (e), 1 (f), 2 (e), 2 (f) and 3 (d).

## 3. Conclusions

We have studied the approximate scattering state solutions of the Klein-Gordon equation with the equal scalar and vector Varshni, Hellmann and Varshni-Shukla potential models using a suitable approximation to overcome the effects of centrifugal terms via functional analytical method. The approximate bosonic scattering phase shifts, wavefunctions, normalization constants and the non-relativistic limit of the scattering the phase shifts for the three potential models were obtained and discussed extensively.

Using the condition or the property at the poles of scattering amplitude, the bound state energy eigenvalues formula for the three physical potential models were obtained. Some of our bound state energy formula are the same with the available one in the literature. For example, the Hellmann potential (see equation (24) of Hamzavi et al. [28] and equation (24) of Onate et al. [30].

To show the beauty and the validity of our work, we obtained the numerical values of the scattering phase shift for different values of potential parameters. The numerical and graphical results in this work show the dependencies of bosonic and non-relativistic scattering phase shifts for the three potential models on the angular momentum number $l$, screening parameter $β$ and potential strengths $b$. The difference and the similarity in the behaviour of the three potentials are seen in their phase shift plots. These results suggest that model parameters should be considered while performing scattering experiment in chemical and nuclear physics.

**Acknowledgements**

Oluwadare O.J declared that this paper represents an output of my Ph.D. research thesis supported by the Tertiary Education Trust Funds (TETFunds) through the Federal University Oye Ekiti, Ekiti State, Nigeria. The authors also declared that there is no conflict of interest regarding this paper.